\documentclass[preprint,superscriptaddress,keywords,preprintnumbers,amsmath,amssymb,aps,prb]{revtex4}
\usepackage{graphicx}
\usepackage{dcolumn}
\usepackage{bm}
\begin{document}
\title{Non-uniform scaling of the magnetic field variations before the $M_w$9.0 Tohoku earthquake in 2011}

\author{E. S. Skordas}
\email{eskordas@phys.uoa.gr}\affiliation{Solid State Section and
Solid Earth Physics Institute, Physics Department, University of
Athens, Panepistimiopolis, Zografos 157 84, Athens, Greece}

\begin{abstract}
Applying Detrended Fluctuation Analysis (DFA) to the geomagnetic data recorded at three measuring
stations in Japan, Rong et al. in 2012  reported that
anomalous magnetic field variations were identified well before
the occurrence of the disastrous Tohoku $M_w$9.0 earthquake that
occurred on 11 March 2011 in Japan exhibiting increased
``non-uniform'' scaling behavior. Here, we provide an explanation
for the appearance of this increase of ``non-uniform'' scaling on
the following grounds: These magnetic field variations are the
ones that accompany the electric field variations termed Seismic
Electric Signals (SES) activity which have been repeatedly
reported that precede major earthquakes. DFA as well as
multifractal DFA reveal that the latter electric field variations
exhibit scaling behavior as shown by analyzing SES activities
observed before major earthquakes in Greece. Hence, when these
variations are superimposed on a background of pseudosinusoidal
trend, their long range correlation properties -quantified by DFA-
are affected resulting in an increase of the ``non-uniform''
scaling behavior. The same is expected to hold for the former
magnetic field variations. 
\end{abstract}

\maketitle

{\bf Complex systems exhibit scale invariant features
characterized by long-range power-law correlations, which are
usually difficult to identify mainly due to the presence of
erratic fluctuations and nonstationarity embedded in the emitted
signals. For example, the long-range correlation properties of the
so called Seismic Electric Signals activities (which are low
frequency electric signals of dichotomous nature preceding major
earthquakes) are affected by pseudosinusoidal trends. By employing
Detrended Fluctuation Analysis (DFA), which has been established
as a robust method suitable for detecting long-range power-law
correlations embedded in non-stationary signals, it is found that
SES activities exhibit infinitely ranged temporal correlations,
i.e., with an exponent close to unity. The same holds for the
magnetic field variations accompanying SES activities. Recently,
the daily scaling properties of geomagnetic field data in the
Japanese area have been investigated by DFA for a period more than
one year before the occurrence of the devastating Tohoku
earthquake of magnitude 9.0 in 2011. Just a few months before its
occurrence, an increase of the ``non-uniform'' scaling behavior
was identified. This is explained as follows: In the absence of
SES activities, the electrical records exhibit a pseudosinusoidal
background due to electric field variations induced by frequent
tiny variations of the Earth's magnetic field of extraterrestrial
origin. Examples of applying DFA to such time series are
presented. Upon approaching a major earthquake, however, the
feature of the electrical records markedly change, because upon
the appearance of the precursory SES activity a large number of
dichotomous electric pulses superimpose on the pseudosinusoidal
background thus leading to significantly different results of DFA
the description of which reveals the existence of three scaling
exponents (``non-uniform'' scaling). A similar behavior is found
when analyzing magnetic records, since magnetic field variations
accompany an SES activity.}

\section{Introduction}
Xu et al. \cite{Xu2013} reported that anomalous variations of the
geomagnetic field have been observed prior to the $M_w$9.0 Tohoku
earthquake (EQ) that occurred in Japan on 11 March 2011. In
particular, the original records of the geomagnetic field at the
Esashi (ESA) station located at about 135 km from the epicenter
exhibited anomalous behavior for about 10 days (4 to 14 January
2011) mainly in the vertical component approximately 2 months
before the $M_w$9.0 earthquake. The findings of \citet{Xu2013}
were found \cite{JAES14} to be in full accord with  results
published independently \cite{TECTO13,PNAS13} on the basis of the
analysis of seismicity of Japan in a new time domain -termed
natural time \cite{NAT02}- which reveals some dynamic features
hidden \cite{ABE05} in the time series of complex systems
\cite{SPRINGER}. Natural time analysis has been applied to diverse
fields including seismicity (e.g., the correlation properties
between consecutive earthquake magnitudes\cite{VAR04,NAT06B} and
how they are affected when a major event is
impending\cite{NAT11B}), heart rate variability\cite{SAR09A} and
systems exhibiting self-organized criticality\cite{NAT11A}. The
results of Refs. \onlinecite{TECTO13,PNAS13} could be summarized
as follows:

First, \citet{TECTO13} identified that the fluctuations of the
order parameter of seismicity defined in natural time
\cite{NAT05C} exhibited a clearly detectable minimum approximately
at the time of the initiation of the pronounced Seismic Electric
Signals (SES) activity observed by \citet{UYE09} almost two months
before the onset of the volcanic-seismic swarm activity in 2000 in
the Izu Island region, Japan. SES are low frequency ($\le 1$Hz)
changes of the electric field of the earth that have been found in
Greece\cite{JAP2,EPL12,NAT02} and Japan\cite{UYE00} to precede
earthquakes with lead times ranging from several hours to a few
months. They are generated by means of the following
mechanism\cite{VARBOOK}: In the focal area there exist ionic
constituents in which the presence of aliovalent impurities
produce extrinsic point defects\cite{VARALEX80} which are attracted
by the impurities thus forming electric dipoles\cite{VARALEX81}. The
relaxation time of these dipoles depend on temperature and
pressure\cite{SPRINGER}. Before an EQ the stress $\sigma$
gradually increases, which may reflect a decrease of the
relaxation time, thus it may reach a {\it critical } value when
$\sigma =\sigma_{cr}$. This results in a cooperative orientation
of the electric dipoles leading to an emission of a transient
electric signal which constitutes an SES. A number of such signals
within a short time is called\cite{JAP2} SES activity and the
natural time analysis of the subsequent seismicity enables the
identification\cite{SAR08,NAT01} of the time-window of the
forthcoming major EQ.

Second, \citet{PNAS13} proceeded to the analysis of the Japan
seismic catalog in natural time from 1 January 1984 to 11 March
2011 and demonstrated that the fluctuations of the order parameter
of seismicity showed distinct minima a few months before all the
shallow earthquakes of magnitude 7.6 or larger during this 27 year
period in Japanese area. Among these minima, the deepest one was
observed before the $M_w$9.0 Tohoku earthquake on $\sim$ 5 January
2011. This fact, in view of the aforementioned findings of
\citet{TECTO13}, reflects that a strong SES activity should have
been initiated on the same date, i.e., on $\sim$5 January 2011.
The SES activities are accompanied by magnetic field variations,
which are clearly detectable at distances of the order of $\sim
100$ km for EQs of magnitude 6.5 or larger \cite{VAR03}, appearing
mainly \cite{SAR02} in the $Z$ component, e.g., see Ref.
\onlinecite{VAR99BB}.

In an independent study \citet{RONG12} analyzed the daily scaling
properties of geomagnetic time series from 1 January 2010 to 30
April 2011 recorded at three stations near the epicenter of the
$M_w$9.0 Tohoku EQ. By employing the detrended fluctuation
analysis (DFA) -see below-, deviations from uniform power-law
scaling were identified and quantified using a scaling index. They
suggested\cite{RONG12} that a significant increase of
``non-uniform'' scaling index appeared well before the Tohoku EQ
and concluded that the scaling properties of the local nonlinear
system are possibly affected by the Tohoku EQ. It is the scope of
this short paper to indicate that this precursory increase of the
``non-uniform'' scaling behavior in the magnetic field data can
find a reasonable explanation  on the basis of the aforementioned
proposal \cite{JAES14}, i.e., that the anomalous magnetic field
variations are the ones which accompany the SES activity initiated
around 5 January 2011 mentioned above.

\section{Detrended Fluctuation Analysis. Monofractals and Multifractals Background}
The signals emitted from complex systems exhibit fluctuations over
multiple scales which are characterized by absence of dynamic
scale, i.e., scale-invariant behavior\cite{STA95}. These signals
are typically non-stationary and their reliable analysis should
not be carried out by traditional methods, e.g., power-spectrum
and auto-correlation analysis\cite{HUR51,MAN69,STR81}. On the
other hand Detrended Fluctuation Analysis (DFA)\cite{PEN94,TAQ95}
has been established as a robust method suitable for detecting
long-range power-law correlations embedded in non-stationary
signals and has been applied with successful results to diverse
fields where scale-invariant behavior emerges, such as
DNA\cite{PEN93}, heart dynamics\cite{IVA99,IVA01}, circadian
rhythms\cite{HU04}, meteorology\cite{IVAN99} and climate
temperature fluctuations\cite{BUN98}, economics\cite{VAN97} as
well as in SES activities\cite{NAT03A,NAT03B,NAT09V} along with
their relevant\cite{EPL12,VAR01B,SAR02,VAR03} magnetic field
variations\cite{NAT09V}.

Monofractal signals are homogeneous in the sense that they have
the same scaling properties, characterized locally by a single
singularity exponent $h_0$, throughout the signal. Thus,
monofractal signals can be indexed by a single global exponent,
e.g., the Hurst\cite{HUR51} exponent $H \equiv h_0$, which
suggests that they are stationary from viewpoint of their local
scaling properties (see Refs. \onlinecite{IVA99,IVA01} and
references therein). Since the traditional DFA can measure only
one exponent, this method is more suitable for the investigation
of monofractal signals (this is alternatively termed as a case of
a ``uniform'' scaling). In this case the root mean square
variability of the detrended process in DFA varies with the scale
$s$ as $F(s)\propto s^{\alpha}$ (see also below). In some cases,
however, the records cannot be accounted for by a single scaling
exponent, i.e., they do not exhibit a simple monofractal behavior.
In general, if a number of scaling exponents is required for a
full description of the scaling behavior, a multifractal analysis
must be applied. A reliable multifractal analysis can be performed
by the Multifractal Detrended Fluctuation Analysis,
MF-DFA\cite{KAN02B} or by the wavelet transform (e.g., see Refs.
\onlinecite{IVA99,MUZ94}).

A brief description of DFA is as follows. We first calculate the
`profile':
\begin{equation}\label{psi}
{y(n)=\sum _{i=1}^n \left( x_i - \overline{x} \right)}
\end{equation}
{of a time series \{$x_i$\}, $i=1,2,...,N$ with mean
$\overline{x}$:}
\begin{equation}\label{mesox}
{\overline{x} = \frac{1}{N} \sum_{i=1}^N x_i}
\end{equation}
where $N$ is the length of the signal.

{Second, the  profile $y(n)$ is divided into $N_s \equiv [N/s]$
non overlapping segments of equal length (``scale'') $s$. Third,
we estimate a (piecewise) polynomial trend $y_s^{(l)}(n)$ within
each segment by least-squares fitting, i.e., $y_s^{(l)}(n)$
consists of concatenated polynomials of order $l$ which are
calculated separately for each of the segments.  The degree of the
polynomial can be varied in order to eliminate linear ($l=1$),
quadratic ($l=2$), or higher order trends\cite{BUN00} of the
profile function. DFA is named after the order of the fitting
polynomial, i.e., DFA-1 if $l=1$, DFA-2 if $l=2$,... Note that,
due to the integration procedure in the first step, DFA-$l$
removes polynomial trends of order $l-1$ in the original signal
\{$x_i$\}. Fourth, the detrended profile function $\tilde{y}_s(n)$
on scale $s$ is determined by}
\begin{equation}\label{mesoy}
{\tilde{y}_s(n)=y(n)-y_s^{(l)}(n)}
\end{equation}
{which, in other words, means that the profile $y(n)$ is detrended
by subtracting the local trend in each segment. Fifth, the
variance of $\tilde{y}_s(n)$ yields the fluctuation function on
scale $s$}
\begin{equation}\label{fluctfun}
{F(s)=\sqrt{\frac{1}{N} \sum_{n=1}^N[\tilde{y}_s(n)]^2}}
\end{equation}
Sixth, the above computation is repeated for a broad number of
scales $s$ to provide a relationship between $F(s)$ and $s$. A
power law relation between $F(s)$ and $s$, i.e.,
\begin{equation}\label{fsmes}
{F(s) \propto s^{\alpha}}
\end{equation}
indicates the presence of scale-invariant (fractal) behavior
embedded in the fluctuations of the signal. The fluctuations can
be characterized by the scaling exponent $\alpha$, a
self-similarity parameter: If $\alpha=0.5$, there are no
correlations in the data and the signal is uncorrelated (white
noise); the case $\alpha < 0.5$ corresponds to anti-correlations,
meaning that large values are most likely to be followed by small
values and vice versa. If $\alpha > 0.5$, there are long range
correlations, which are stronger\cite{BAS08} for higher $\alpha$.
Note that $\alpha > 1$ indicates a non-stationary local average of
the data and the value $\alpha=1.5$ indicates Brownian motion
(integrated white noise).

Compared to DFA in the MF-DFA the following additional two steps
should be made:

First, we average over all segments to obtain the $q-$th order
fluctuation function $F_q(s)$:

\begin{equation}\label{(13.4.1.3)}
F_q(s)\equiv \left\{ \frac{1}{N_s} \sum_{\nu=1}^{N_s} \left[
F^2(s,\nu) \right]^{\frac{q}{2}} \right\}^{\frac{1}{q}}
\end{equation}
where
\begin{equation}
F^2(s,\nu)=\frac{1}{s} \sum_{n=(\nu-1) s+1}^{\nu s}
\tilde{y}_s(n)^2 ,
\end{equation}
and the index variable $q$ can take any real value except zero.
This is repeated for several scales $s$.

Second, we determine the scaling behavior of the fluctuation
functions by analyzing log-log plots $F_q(s)$ versus $s$ for each
value of $q$. For long-range  power-law correlated series,
$F_q(s)$ varies as
\begin{equation}\label{(13.4.1.4)}
F_q(s)\propto s^{h(q)},
\end{equation}
where the function $h(q)$  is called generalized Hurst exponent.
For stationary time series the aforementioned Hurst exponent $H$
is identical to $h(2)$,
\begin{equation}\label{(13.4.1.5)}
h(2)=H.
\end{equation}

\section{Results of Applying DFA to SES activities and their associated magnetic field variations}
When trying to identify anomalous electric field variations, and
hence SES activities, our electrical records are contaminated by
noise mainly due to the following two origins: First, artificial
noise (AN) comprising electric signals emitted from man-made
sources located close to the measuring station. AN may look to be
similar to SES activities. Second, electric variations induced by
small changes of the Earth's magnetic field (magnetotelluric
variations, MT) of extraterrestrial origin are practically
continuously superimposed on our records. MT variations are easily
recognized when a network of measuring stations is operating since
they appear almost simultaneously at the records at all measuring
sites. On the other hand, the recognition of AN is more difficult
and can be achieved by means of DFA and MF-DFA upon employing also
natural time analysis as follows:

When DFA is applied to the original time series of the SES
activities and AN, it was found\cite{NAT03A} that both types of
signals lead to a slope at short time scales (i.e., $\Delta t \leq
30s$) lying in the range $\alpha$=1.1-1.4, while for longer time
scales the range $\alpha$=0.8-1.0 was determined without, however,
any safe classification between SES activities and AN. On the
other hand, when employing natural time, DFA enables the
distinction between SES activities and artificial noises: for the
SES activities the $\alpha$-values lie approximately in the range
0.9 - 1.0, while for the AN the $\alpha$-values are markedly
smaller, i.e., $\alpha$=0.65-0.8. This reflects that the SES
activities exhibit infinitely ranged temporal correlations, which
is not the case for the AN. The fact that $\alpha \approx 1$ has
been also verified \cite{NAT09V} for the magnetic field variations
accompanying the SES activities.

MF-DFA was also applied\cite{NAT03A} to the time series of SES
activities and AN. This multifractal analysis, when carried out in
the conventional time frame, did not lead to any distinction
between these two types of signals. On the other hand, if the
analysis is made in natural time, a distinction becomes
possible\cite{NAT03A}. In particular, when the MF-DFA is applied
in natural time reveals that the $h(q)$ curves for the SES
activities lie systematically higher than those in the case of
``artificial'' noises. For example, for $q=2$ the $h(2)$ values
for the SES activities lie close to unity. A similar behavior is
expected to hold also for the magnetic field variations associated
with the SES activities. On the other hand, the $h(2)$ values for
AN scatter \cite{NAT03B} approximately in the range 0.65-0.8  (see
also Subsection 4.5.3 of Ref.\onlinecite{SPRINGER}).

Multifractal analysis of various SES activities in Greece has been
also performed \cite{NAT03B} by using the wavelet transform
instead of MF-DFA.

\begin{figure}
\includegraphics[scale=0.5]{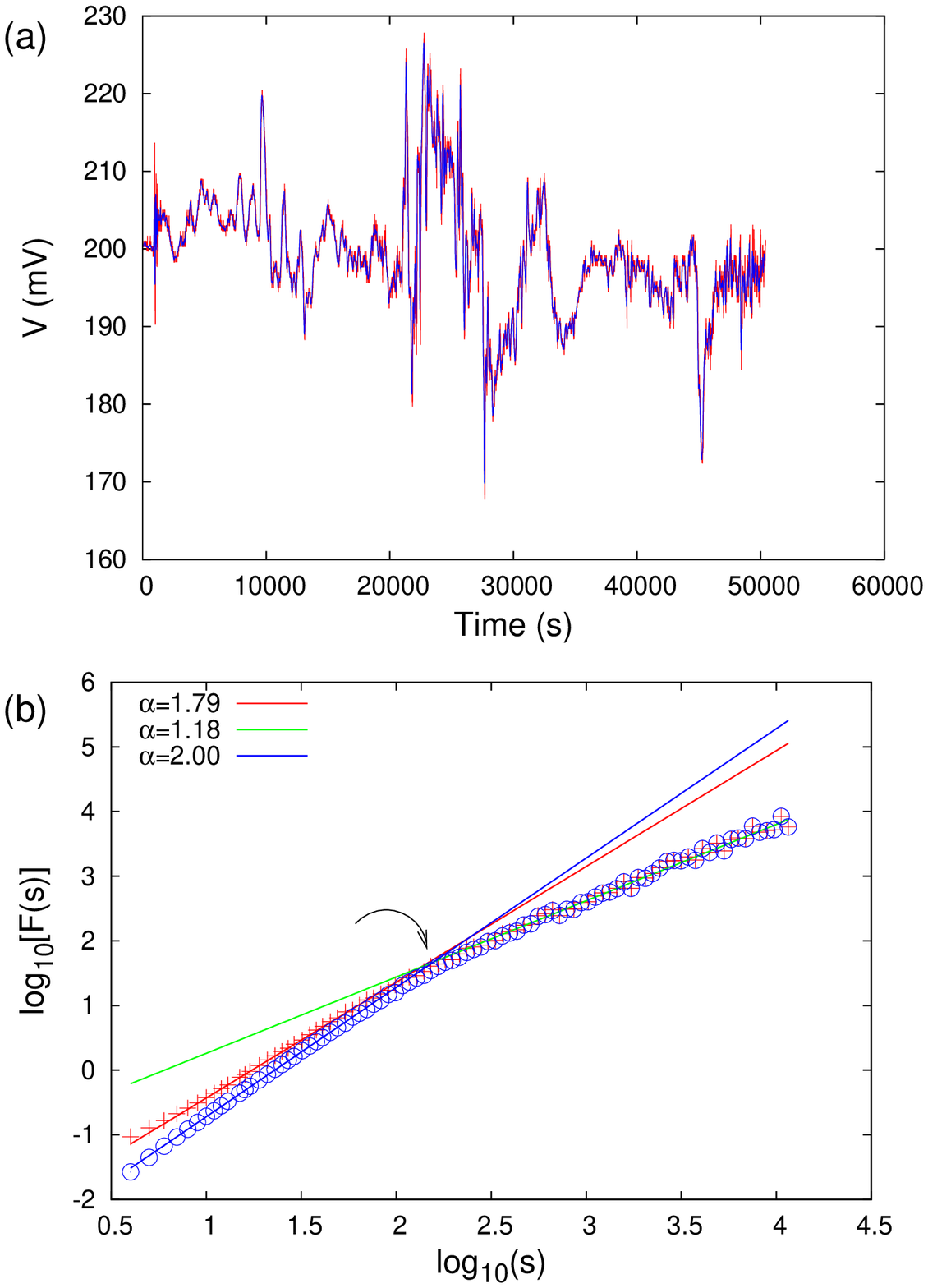}
\caption{(color online) (a) An almost 14h excerpt of the
electrical field records during which MT variations were also
present on 7 December 2013 at a station located close to Patras
city in Western Greece; red: as received when using a sampling
frequency 1 sample/sec; blue: when plotting (at each point) the
average value of its previous 60 measurements (b) the DFA-1 plots
for the time series mentioned in (a) upon using the same colors.
There exists a single cross-over indicated by the arrow.}
\label{fig1}
\end{figure}

In view of the above, it is thereafter taken as granted that AN
are distinguished from SES activities. Hence, in what remains we
focus on the study of the case of electrical records when MT are
present together with the SES activities by employing DFA. To
facilitate our study we first start from an example in which SES
activities are absent and hence only MT are superimposed on our
records. Such a recent example is depicted in Fig.\ref{fig1}(a)
that has been recorded on 7 December 2013 at a station located
close to Patras city in Western Greece. The red color corresponds
to the record as received of $\sim$14  hour duration taken with
sampling frequency 1 sample/sec while in the blue one in each
point we have plotted the average values of the previous 60
measurements (in a similar fashion as \citet{RONG12} did). The DFA
plots resulting from the analysis of both these curves are shown
in Fig.\ref{fig1}(b) by using the corresponding color. An
inspection of this figure reveals that in both cases there exists
a single crossover (marked with an arrow) at the time scale
$\sim$160s. At the part of the DFA plot corresponding to shorter
scales we deduce an $\alpha$-exponent around $\alpha \sim 1.8$ for
the former curve and $\alpha \sim 2.0$ for the latter while at the
larger scales the slope of this log-log plot for both curves leads
to a smaller exponent, i.e., $\alpha \sim 1.2$. Our finding that
in the shorter scales the $\alpha$-exponent is $\alpha \sim 2.0$
agrees with the results of Fig. 5 of \citet{HU01} when they
studied pseudosinusoidal functions with different amplitudes and
periods.

\begin{figure}
\includegraphics[scale=0.5]{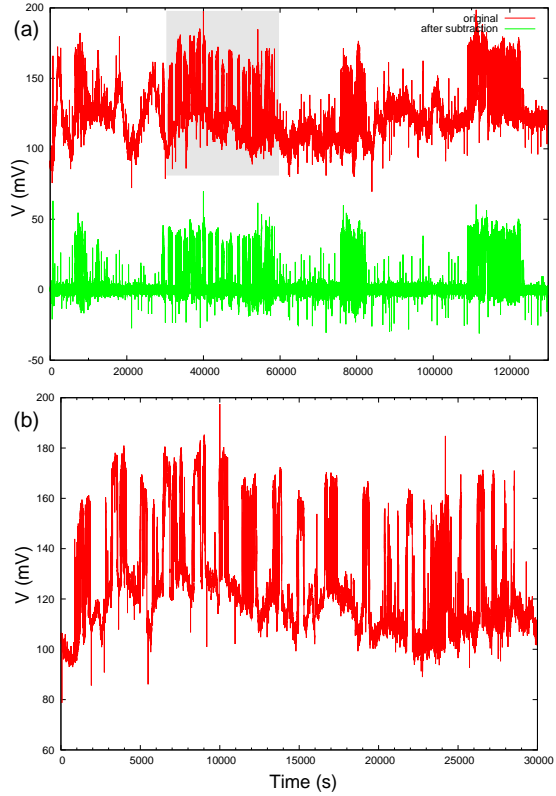}
\caption{(color online) (a) The red color represents the original
record of the SES activity at a station close to Pirgos town in
Western Greece, which lasted from 29 February to 2 March 2008.
This SES activity was superimposed on a pseudosinusoidal
background, due to MT variations, which are evident in the left
part of the figure. After subtracting the MT background -by means
of the procedure described in Ref. \onlinecite{NAT09V}- the green
signal of dichotomous nature remains which constitutes the true
precursory signal, i.e., the SES activity. (b) An almost 8.3 h
excerpt of the original record (shaded in (a)) is plotted here in
expanded time scale.} \label{fig2}
\end{figure}

\begin{figure}
\includegraphics[scale=0.5]{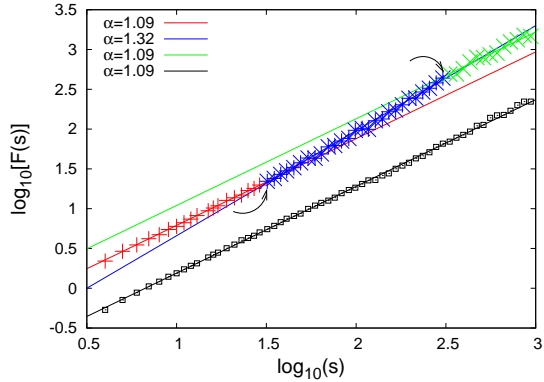}
\caption{(color online) The DFA plot resulting from the analysis
of an excerpt -shaded in Fig. \ref{fig2}(a)- of the original time
series of the SES activity plotted in red in Fig. \ref{fig2}(a)
superimposed on MT variations. It comprises three parts -shown in
red, blue and green- two of which, i.e., the ones corresponding to
the shortest and the largest scales, are depicted in red and
green, respectively. The two cross-overs observed are marked with
arrows. For the sake of comparison, the plot in black depicts the
results deduced from the DFA analysis (shifted vertically for the
sake of clarity) when considering the SES activity alone, i.e.,
the signal of dichotomous nature plotted in green in Fig.
\ref{fig2}(a)} \label{fig3}
\end{figure}

We now turn to a second example in which, beyond MT variations,
SES activities are present in our electrical records. This, in
other words, means that we have considered a time period close to
the occurrence of an impending strong EQ. Such an example is
depicted in Fig.\ref{fig2}(a) in which the red color presents the
original recording of an SES activity collected at a station
located close to Pirgos town in Western Greece from 29 February to
2 March 2008. This SES activity, which preceded the disastrous
$M_w$6.5 EQ that occurred with an epicenter at $38.0^o$N $21.5^o$E
lying between Pirgos and Patras only a few tens of km away from
the measuring site, is superimposed on a pseudosinusoidal
background arising from MT variations. The procedure through which
this pseudosinusoidal background can be subtracted should also
employ natural time analysis and has been described in detail by
\citet{NAT09V}. After subtracting it, we find the channel shown in
green which is the true SES activity of obvious dichotomous nature
(this is the lowest channel ``e'' of Fig. 4 of Ref.
\onlinecite{NAT09V}). We now apply DFA to an excerpt of the
original record -shaded in Fig.\ref{fig2}(a)- of several hours
duration ($\sim 8.3$ h) which contains a portion of the SES
activity. This excerpt is shown in expanded time scale in
Fig.\ref{fig2}(b). The DFA plot resulting from this excerpt is
depicted in Fig.\ref{fig3} and comprises three parts shown in red,
blue and green. Two cross-overs (marked with arrows) are observed
at the time scales $\sim$32s and $\sim$320s. The parts
corresponding to the shortest and the largest scales (i.e., the
ones shown in red and green, respectively) lead to a more or less
similar slope, i.e., $\alpha \sim 1.1$, which is very close to
unity, while the intermediate part result in $\alpha \sim 1.3$. In
other words, when comparing the DFA plots depicted in Fig.
\ref{fig1}(b) and Fig. \ref{fig3} exhibiting one cross-over and
two cross-overs respectively, which could be interpreted it as
showing an increase of the ``non-uniform'' scaling behavior upon
analyzing an excerpt of the original record that contains MT
variations (pseudosinusoidal) together with a portion of the SES
activity (dichotomous nature, see also below). By the same token,
a similar behavior should be also found when analyzing the
magnetic field variations -instead of the electric field ones-
accompanying an SES activity. For the sake of comparison, we also
plot (in black) in Fig.\ref{fig3} the results deduced from the DFA
analysis of the SES activity alone, i.e., the signal of
dichotomous nature plotted in green in Fig.\ref{fig2}(a). This
reveals an almost linear $log F(s)$ vs $log s$ plot with an
exponent $\alpha \approx 1$ which remains practically the same
irrespective if we apply DFA-1, DFA-2 or DFA-3, see Ref.
\onlinecite{NAT09V}.

\begin{figure}
\includegraphics[scale=0.5]{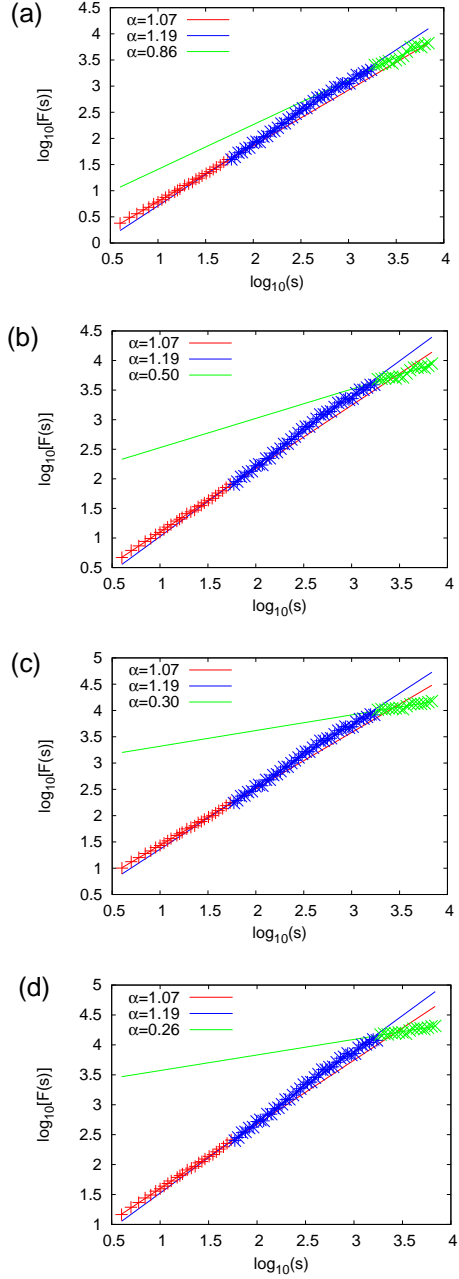}
\caption{(color online) The DFA plots resulting from the analyses
of excerpts of SES activities that precede EQs of various
magnitudes $M_w$=6.5 (a), 7.5 (b), 8.5 (c) and 9.0 (d) (see the
text)} \label{fig4}
\end{figure}

To examine what happens on the ``non-uniform'' scaling behavior
when the amplitude of the SES activity increases we work as
follows:

The extent to which the ``non-uniform'' scaling behavior depends
on the amplitude $E$ of the SES activity can be visualized in Fig.
\ref{fig4}. In particular, the upper panel of Fig. \ref{fig4}
depicts the DFA plot of the 8.3 hour excerpt -shaded in Fig.
\ref{fig2} (a) and shown in expanded time scale in \ref{fig2} (b)-
containing a portion of the SES activity that preceded the
aforementioned $M_w$6.5 EQ in Greece in 2008. Based on this
portion and by making use of the scaling relation $log E = 0.33 M
+$ const that interconnects the SES amplitude with the magnitude
$M$ of the subsequent EQ, we constructed the corresponding
portions that would appear before EQs of stronger magnitude and
analyzed then by DFA. Specifically, the other panels b, c, d of
Fig. \ref{fig4} refer to the DFA plots resulting from the analysis
of SES activities of larger amplitude that correspond to stronger
EQs of magnitude $M$= 7.5, 8.5 and 9.0, respectively. An
inspection of these plots reveals that a systematic change
emerged, i.e., the exponent $\alpha$ at the largest scales
decreases upon increasing the EQ magnitude; the $\alpha$ values,
as shown in Fig. \ref{fig4}, are $\alpha$=0.86, 0.50, 0.30 and
0.26 for $M+w$=6.5, 7.5, 8.5 and 9.0, respectively. In other
words, the ``non-uniform'' scaling seems to be more evident when
analyzing SES activities of dichotomous nature that precede EQs of
larger magnitude.

\section{Summary and Conclusions}
Here, we showed that the  DFA plot resulting from a time-series of
a regular electrical record in which only MT variations are
superimposed, it exhibits a single cross-over. On the other hand,
when our analysis refers to a period before a major EQ and hence
the electrical record contains also an excerpt of an SES activity,
the DFA plot reveals an evident increase of the ``non-uniform''
scaling behavior since two cross-overs now emerge. This should
also hold for the corresponding magnetic field records, i.e., when
an excerpt of the magnetic field variations accompanying an SES
activity is contained in the data analyzed. This explains the
increase of the ``non-uniform'' scaling behavior observed by
\citet{RONG12}, since -as we suggested in Ref.
\onlinecite{JAES14}- the anomalous magnetic field variations
identified by \citet{Xu2013} well before the 2011 $M_w$9.0 Tohoku
EQ might be the ones that accompany a strong SES activity on
$\sim$5 January 2011.


\pagebreak

\end{document}